\documentclass{article}
\usepackage{amssymb}
\usepackage{amsfonts}
\usepackage{amsmath}

\input{tcilatex}

\begin{document}

\title{Bose-Einstein condensation of a Knudsen gas}
\author{Kerson Huang \\
Physics Department, Massachusetts Institute of Technology\\
Cambridge, MA 02319\\
Email: kerson@mit.edu}
\maketitle

\begin{abstract}
We reconcile a long-standing controversy regarding the transition
temperature of the Bose-Einstein condensation in a dilute interacting Bose
gas, by showing that there is a crossover between ideal gas and interacting
gas. The former corresponds to a Knudsen (or collisionless) regime, in which
the mean-free-path is much larger than the system dimension, while the
latter corresponds to the opposite hydrodynamic regime. The deviation of the
transition temperature from that of the non-interacting gas is proportional
to $\sqrt{a}$ in the Knudsen regime, and $a$ in the hydrodynamic regime,
where $a$ is the scattering length. This crossover may be observable in a
Bose gas trapped in a potential or on an optical lattice.
\end{abstract}

\bigskip

The transition temperature $T_{c}$ of a dilute interacting Bose gas in 3D
has been a subject of much theoretical interest and controversy. Let us
denote its fractional deviation from that of the non-interacting gas by

\begin{equation}
\Delta \equiv \frac{T_{c}-T_{c}^{(0)}}{T_{c}^{(0)}},
\end{equation}%
where $T_{c}^{(0)}=\frac{2\pi \hbar ^{2}}{mk_{B}}\left[ n\zeta \left(
3/2\right) \right] ^{-2/3}$, $\zeta \left( 3/2\right) \approx 2.612$, $%
n=N/L^{3},$ with $N$ \ the number of particles and $L$ the linear size of
the system. Using an argument based on the virial expansion in the
high-temperature phase, the author \cite{hua99} obtained the following
result:%
\begin{eqnarray}
\Delta _{0} &=&c_{0}\sqrt{an^{1/3}},  \notag \\
c_{0} &=&\frac{8\sqrt{2\pi }}{3\left[ \zeta \left( 3/2\right) \right] ^{2/3}}%
\approx 3.527.  \label{zero}
\end{eqnarray}%
where $a$ is the S-wave scattering length. This has been rederived in a
mean-field approach \cite{kim07}. \ Earlier, Toyada \cite{toy82} got the
same result but with an opposite sign, by summing one-loop graphs. Many
other calculations, which approach the transition point from the
low-temperature side, obtain a different behavior, with a linear dependence
on $a$ \cite{bij96}--\cite{hol01}:%
\begin{equation}
\Delta _{1}=c_{1}an^{1/3}\text{.}  \label{one}
\end{equation}%
The constant $c_{1}$ is of order 1.

The purpose of this note is to point out that there is a crossover between
ideal gas and interacting gas. The former corresponds to a Knudsen (or
collisionless) regime, in which the mean-free-path in the condensate is much
larger than the size of the system, while the latter corresponds to the
opposite hydrodynamic regime. The fractions $\Delta _{0}$ and $\Delta _{1}$
apply in the Knudsen and hydrodynamic regime, respectively.

Eq.(\ref{zero}) for $\Delta _{0}$ results from a straightforward
generalization of the method used in the ideal gas, namely, by calculating
the maximum number of particles the gas phase can accomodate. That is, one
locates the onset of condensation in the gas phase. The value (\ref{one})
for $\Delta _{1}$, on the other hand, is obtained by approaching the
transition from below, within the condensed phase. These approaches lead to
different answers because of a peculiar property of the perturbation theory
of Bose-condensed systems, namely, the expansion parameter crosses over from 
$a$ at extremely low densities to a fractional power of $a$ at finite
densities.

The ground-state energy per particle of a uniform dilute Bose gas in
conventional perturbation theory is, to second order in $a$, given by \cite%
{hua57} \cite{hua58} 
\begin{equation}
\frac{E_{0}}{\left( \hbar ^{2}/2m\right) N}=4\pi an\left( 1+A_{2}\frac{a}{L}%
\right)
\end{equation}%
where \cite{lus86} \cite{yan87}%
\begin{equation}
A_{2}=2.837297\cdots .
\end{equation}%
In the same expansion scheme, the excitation spectrum has an energy gap $%
8\pi an\xi $, where $\xi $ is the condensate fraction, and there are no
phonons. This expansion breaks down in the thermodynamic limit $N\rightarrow
\infty $, $L\rightarrow \infty $, with $n=N/L^{3}$ held fixed. While the
correction term $a/L$ vanishes, one finds that higher-order corrections
diverge.

A systematic graphical analysis \cite{hua58} reveals that the dimensionless
expansion parameter is 
\begin{equation}
\frac{aN}{L}\thicksim \sqrt{\frac{\text{System size}}{\text{Mean-free-path
in condensate}}}.
\end{equation}%
The mean-free-path in the condensate is \ of order $\left( n\sigma \right)
^{-1}$, with scattering cross section $\sigma \thicksim $ $Na^{2}$, where
the factor $N$ comes from Bose enhancement. The perturbation series has the
general structure

\begin{eqnarray}
\frac{E_{0}}{\left( \hbar ^{2}/2m\right) N}=4\pi an &+&\frac{1}{NL^{2}}\left[
A_{2}\left( \frac{aN}{L}\right) ^{2}+A_{3}\left( \frac{aN}{L}\right)
^{3}+\cdots \right.  \notag \\
&&+\frac{B_{2}}{N}\left( \frac{aN}{L}\right) ^{2}+\frac{B_{3}}{N}\left( 
\frac{aN}{L}\right) ^{3}+\cdots  \notag \\
&&\left. +\frac{C_{2}}{N^{2}}\left( \frac{aN}{L}\right) ^{2}+\frac{C_{3}}{%
N^{2}}\left( \frac{aN}{L}\right) ^{3}+\cdots \right]  \label{expand0}
\end{eqnarray}%
This expansion obviously fails in the thermodynamic limit, where $aN/L$
diverges. To obtain a finite energy per particle, one reorganizes the
perturbation series by effectively performing infinite sums in each
horizontal line in (\ref{expand0}). One can see on general grounds that the
first horizontal sum should be proportional to $a^{5/2}$. Explicit
calculation gives an expansion in powers of $\sqrt{na^{3}}$ in the
thermodynamic limit \cite{lee57}:%
\begin{equation}
\frac{E_{0}}{\left( \hbar ^{2}/2m\right) N}\longrightarrow 4\pi an\left[ 1+%
\frac{128}{15\sqrt{\pi }}\sqrt{na^{3}}+O\left( na^{3}\right) \right] .
\label{expand1}
\end{equation}%
The excited spectrum now consists of phonons, with no energy gap. The
expansion scheme (\ref{expand0}), which is valid for the $N$-body system
with fixed $N$ in an infinite volume, yields the virial series for the
equation of state \cite{hua57b}, and that leads to $\Delta _{0}$ given by (%
\ref{zero}). The partial summation of the perturbation series leading to (%
\ref{expand1}) is equivalent to the Bogoliubov transformation, with the
emergence of \textquotedblleft pairing", and long-wavelength collective
modes that contribute significantly to $\Delta _{1}$ in (\ref{one}).

The Knudsen regime is relevant for kinetic processes, such as effusion
through a hole, when the mean-free-path is much larger than the dimension $L$
of the hole. For a macroscopic system in thermal equilibrium,\ $L$ is the
size of the whole system, and the Knudsen region consists of an
infinitesimally small neighborhood of zero density, and is of little
experimental interest.

For a Bose gas\ condensed in a potential or on an optical lattice, the
Knudsen regime spans a region accessible to experimentation, for one can
vary the system size, particle number, and even the scattering length
through Feshbach resonances. In a harmonic trap of frequency $\omega _{0}$,
the relevant dimension is the harmonic length%
\begin{equation}
L=\sqrt{\frac{\hbar }{m\omega _{0}}}.
\end{equation}%
The trapped gas behaves like an ideal gas when $aN/L\ll 1$, and crosses over
to the Thomas-Fermi regime when $aN/L\gg 1$. During the crossover, the size
of the condensate increases from the ideal-gas radius%
\begin{equation}
R_{0}\thicksim L
\end{equation}%
to the Thomas-Fermi radius \cite{tim75}%
\begin{equation}
R_{\text{TF}}\thicksim \left( \frac{aN}{L}\right) ^{1/5}L.
\end{equation}%
The excitations change from ideal-gas states in the trap to collective
oscillations, the lowest ones being the \textquotedblleft breathing"
(monopole) and quadrupole modes \cite{dal99}. Note that the ideal gas regime
is not the same as that of no interaction. Even a very weak interaction
leads to quantum phase coherence that is absent in the non-interacting
system.

As illustration, take $a=1$ nm, $L=100$ $\mu $m$.$The critical particle
number is%
\begin{equation}
N_{c}=\frac{L}{a}=10^{5}.
\end{equation}%
The transition temperature depends on $N$, and interpolates between two
fractional deviations:%
\begin{equation}
\Delta (N)=\left\{ 
\begin{array}{cc}
\Delta _{0} & \left( N\ll N_{c}\right) \\ 
\Delta _{1} & \left( N\gg N_{c}\right)%
\end{array}%
\right. .
\end{equation}%
The values given in (\ref{zero}) and (\ref{one})\ are for the uniform gas,
but may be used for order-of-magnitude estimates. The former corresponds to
a 1\% correction, while the latter amount to 0.02\%. The Knudsen gas has a
higher transition temperature, because it is harder to excite the condensate
in that case, owing to the energy gap.

\end{document}